\documentclass[fleqn,usenatbib]{mnras}
\usepackage{newtxtext,newtxmath}
\usepackage[T1]{fontenc}
\usepackage{ae,aecompl}
\usepackage{graphicx}
\usepackage{amsmath}
\usepackage{amssymb}	
\title[Interacting Young M-dwarfs]{Interacting Young M-dwarfs in Triple System - Par 1802 Binary System Case Study}
\author[Cheng et al.]{Shelley J. Cheng,$^{1}$\thanks{E-mail: shelleycheng@ucla.edu}
Alec M. Vinson,$^{1}$
Smadar Naoz$^{1,2}$
\\
$^{1}$Department of Physics and Astronomy, University of California, Los Angeles, CA 90095, USA\\
$^{2}$Mani L. Bhaumik Institute for Theoretical Physics, Department of Physics and Astronomy, UCLA, Los Angeles, CA 90095, USA
}
\date{Accepted XXX. Received YYY; in original form ZZZ}
\pubyear{2019}
\begin{document}
\newpage
\label{firstpage}
\pagerange{\pageref{firstpage}--\pageref{lastpage}}
\maketitle
\begin{abstract}
The binary star Par 1802 in the Orion Nebula presents an interesting puzzle in the field of stellar dynamics and evolution. Binary systems such as Par 1802 are thought to form from the same natal material and thus the stellar members are expected to have very similar physical attributes. However, Par 1802's stars have significantly different temperatures despite their identical (within $3\%$) masses of about $0.39$ solar mass. The leading proof-of-concept idea is that a third companion gravitationally induced the two stars to orbit closer than their Roche-limit, which facilitated heating through tidal effects. Here we expand on this idea and study the three-body dynamical evolution of such a system, including tidal and pre-main-sequence evolution. We also include tidal heating and mass transfer at the onset of Roche-limit Crossing. We show, as a proof-of-concept, that mass transfer combined with tidal heating can naturally explain the observed temperature discrepancy. We also predict the orbital configuration of the possible tertiary companion. Finally we suggest that the dynamical evolution of such a system has pervasive consequences. We expect an abundance of systems to undergo mass transfer during their pre-main-sequence time, which can cause temperature differences. 
\end{abstract}
\begin{keywords}
stars: kinematics and dynamics -- stars: pre-main-sequence --  binaries: general -- binaries: close
\end{keywords}

\section{Introduction}
Basic physical properties of a star such as mass and temperature are strongly determined by its initial mass and chemical composition. Equal-mass components of binary stars are typically considered to be coeval and identical `twins' formed from the same protostellar core, and are expected to have identical physical attributes. However, a small number of identical twin systems seem to be inconsistent with this narrative, featuring a large temperature difference between its component stars. These systems are rare, with $\sim 5$ such systems having been observed and studied \citep[e.g.,][]{Cargile+08,Gomez+09,Gomez+19,Gillen+2017,Wang+2009}. We suggest that this type of system is a natural result of dynamical evolution in three-body systems.

As a case study, we focus on the Par 1802 binary system \citep{Stassun+08}. This system is a relatively young \citep[few-million-year-old, see][]{Cargile+08,Gomez+12,Stassun+14} pre-main-sequence `twin' composed of two M-dwarf stars, each with identical (to within $3\%$) masses of $0.391$~M$_\odot$ and $0.385$~M$_\odot$ \citep{Gomez+12,Stassun+14}. However, the temperatures of the binary stars are $3675$~K and $3365$~K, which surprisingly differ by $\sim300$~K (or $9.2\%$) \citep{Gomez+12,Stassun+14} despite their otherwise identical physical attributes. 

The majority \citep[$\sim 75\%$][]{Salpeter+55,Henry+06} of all the stars in our Galaxy are low mass (about $10\%-50\%$ the mass of the Sun) and cold (about $30\%-65\%$ the temperature of the Sun) stars that belong to the M-dwarf (red dwarf) class. Approximately $20-40\%$ of all M-dwarfs have a binary companion \citep{Janson+12, Bergfors+10, Fischer+92,Raghavan+10}, and in a survey of $151$ contact binaries using data from the \textit{Hipparcos} satellite mission, \citet{Pri+06} found that $42\% \pm 5\%$ are at least in triple systems. Therefore, many binary stars in our galaxy are in fact in triple configurations \citep{T97, Tok+06, Egg+07,Griffin2012,Rappaport+13}.

The prevalence of M-dwarf binaries in triple systems motivated \citet{Gomez+12} and \citet{Stassun+14} to study Par 1802 in the presence of a third star. In particular, they proposed that the temperature difference between the components of Par 1802 is likely due to different spin evolution and uneven mass accretion, perhaps caused by pre-main-sequence tidal evolution between the two stars. However, their work did not include the dynamical evolution of the three body system, which can dramatically alter the binary's orbital parameters \citep[e.g.][]{Naoz16}. Here we follow the dynamical evolution of a Par 1802-like system and show that a distant tertiary can naturally explain all of the observed properties. Moreover we constrain the orbital configuration of the third companion. 

Dynamical stability of such a system requires a hierarchical configuration where a third body on a far wider (outer) orbit orbits around the tighter (inner) binary of Par 1802. In this configuration, the triple body orbital parameters change on secular timescales. In particular, gravitational perturbations from the tertiary star can cause eccentricity excitations of the inner orbit, i.e., Kozai-Lidov cycles \citep{Kozai, Lidov, Naoz16}. The two non-resonant orbits can be described by two ``wires'' where the line-densities are inversely proportional to orbital velocity. Gravitational potential can then be expanded in the semi-major axis ratio, as the hierarchical configuration drives the ratio between the inner and outer body's semi-major axes to a small parameter. Here, we employ the hierarchical three-body secular approximation up to the octupole-level of approximation, known as the Eccentric Kozai Lidov (EKL) mechanism \citep[e.g.][]{Naoz16}. It was recently shown that the EKL mechanism plays a pivotal role in the evolution of triple stellar systems \citep{Thompson2011, Naoz+13sec, NF, Naoz16, Stephan+16,Bataille+18,Moe+2018}. Gravitational perturbations from a third companion excite high eccentricities in the inner binary, thus the stars spend longer times near each other, yielding to stronger tides that can shrink the binary separation and circularising it. This process is very effective in forming short period inner binaries such as Par 1802 \citep[e.g.][]{NF}.

This paper provides a theoretical background (Section~\ref{sec:theory}), describes the setup for the dynamical simulations (Section~\ref{sec:methodology}), presents and analyses results of the simulations (Section~\ref{sec:results}) with particular emphasis on predictions of the third companion's orbital parameters. We investigate the interaction at Roche-limit and explain the observed temperature difference between the stars of the inner binary (Section~\ref{sec:interaction}) through a proof-of-concept mass transfer analysis and a tidal heating model. A discussion of our results is offered in Section~\ref{sec:discussion} before drawing our conclusions.

\section{Theory}\label{sec:theory}
\subsection{Pre-main-sequence Evolution }
\begin{figure}
	\includegraphics[width=\columnwidth]{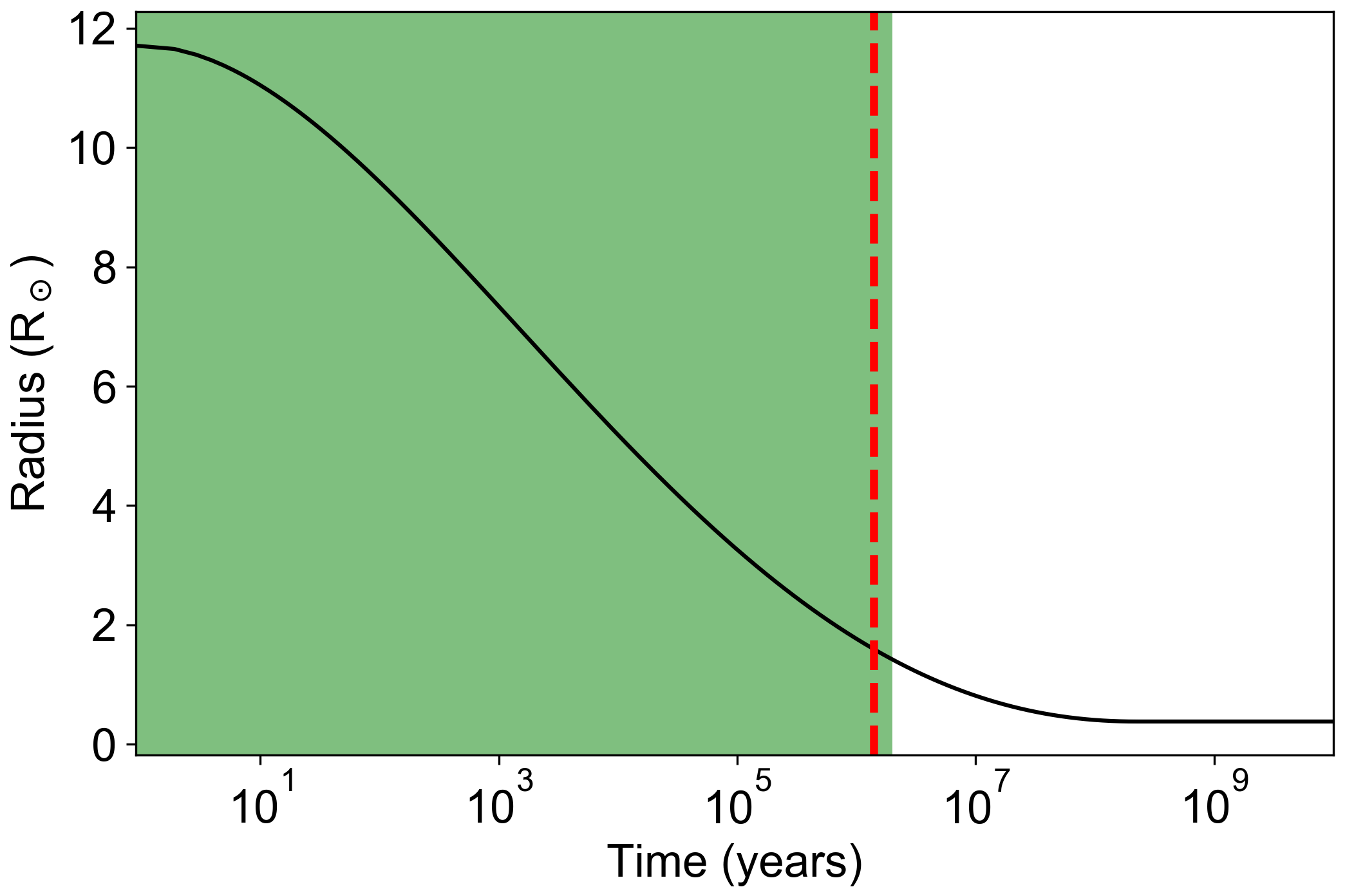}
	\caption{\textbf{Pre-main-sequence radius contraction for a $\mathbf{0.391}$~M$_\odot$ M-dwarf star.} Interpolated from \citet{Chabrier&Baraffe97}, with pre-main-sequence (and thus radius contraction) ending at $2.1 \times 10^8$~years. The age of the system of $\sim 2$~Myrs \citep{Stassun+14, Gomez+12} is shaded in green. The estimated Roche-limit crossing time of $1.4$~Myrs is in red.}
	\label{fig:pms}
\end{figure}

Tidal evolution of such a compact binary system is very sensitive not only to orbital separation, but also to stellar radii. As we are modelling the tidal evolution during the stars' pre-main-sequence lifetime, it is therefore very important to also consider the dramatic evolution in stellar radii during this time. To do this, we consider the calculations of \cite{Chabrier&Baraffe97} for the radius evolution of small ($< 0.8\ M_\odot$) pre-main-sequence stars, wherein they consider stellar evolution as a function of mass and metallicity using grainless non-grey atmosphere models.  We interpolate their results, assuming solar metallicity, for pre-main-sequence stars to find radius for each star as a function of its mass and of time while modelling the tidal evolution. The radii of these stars dramatically contracts during their pre-main-sequence lifetime, with initial inflated radii of approximately $11$ times the radius of the Sun, to final radii of approximately $1.5$ times the radius of the Sun (see Figure~\ref{fig:pms}). This in turn has dramatic effects on the tidal evolution, with the tidal forces weakening greatly as the star contracts.

\subsection{Analytical Expectations for the Tertiary companion}
Par 1802 is a very tight binary with semi-major axis $a\sim 0.049$~au and on a circularised orbit of eccentricity $e\sim0.02$  \citep{Gomez+12}. The temperature difference between the stars implies that mass transfer may have taken place during Roche-limit crossing phase of the evolution. The Roche-limit, $R_{\rm Roche, 1}$ of a body of mass $m_1$ and radius $R_1$ with respect to an orbiting body of mass $m_2$ is defined as:
\begin{equation}
R_{\rm Roche, 1} = 2.7 \times R_1 \left(\frac{m_1 + m_2}{m_1}\right)^{-{1}/{3}} \ .
\label{eq:Roche}
\end{equation}
The numerical pre-factor is highly uncertain and here we adopt $2.7$ following numerical simulations from \citet{Guillochon+11} and \citet{Liu+13}. As the M-dwarf goes through pre-main-sequence contraction the Roche-limit contracts. Thus, a system may have a short phase of mass transfer during the evolution before settling down on a stable tight configuration.  

The two stars probably began their life on a wider orbit than currently observed, avoiding merging at birth. The tertiary induces large values of eccentricity that eventually, with tidal evolution, drive the binaries to a tight configuration \citep[e.g.,][]{Egg1998,Kiseleva1998,Eggleton2001,FabryckyTremaine07,PeretsFabrycky09,Thompson2011,Shappee+13,NF}. The final semi-major axis from this process is most likely the result of tidal shrinking that conserved angular momentum \citep[e.g.,][]{Ford+2000}. In other-words, we can write:
\begin{equation}
    a_i(1-e_i^2) \approx a_f \ ,
\end{equation}
where subscript ``$i$'' refers to initial configuration (before tidal effects) and subscript ``$f$'' refers to final configuration, $a$ is the semi-major axis, and $e$ is the eccentricity.
High eccentricity migration implies that $e\to 1$ during the evolution, thus,
\begin{equation}\label{eq:af}
    a_f\sim 2 a_i(1-e_i)\sim 2 R_{\rm Roche} \ ,
\end{equation}
assuming that the closest approach before disruption can be the Roche-limit. In our case, where mass transfer may have taken place, $R_{\rm Roche}$ represents the maximum approach for interaction.

Adopting the observed semi-major axis of Par 1802 $a_f = 0.049$~au, and using Equation~(\ref{eq:af}), we find that the pericentre crosses the Roche-limit at a radius:
\begin{equation}
    R_1(t) \sim \frac{a_f}{2 \times 2.7} \left(\frac{m_1 + m_2}{m_1}\right)^{-1/3} \sim 1.57~R_\odot \ .
\end{equation}
We adopt pre-main-sequence evolution of M-dwarfs following  \citet{Chabrier&Baraffe97}, and can estimate the time the star reached this value (the vertical line in Figure \ref{fig:pms}). We thus estimate that the limit on the Roche-limit crossing time was at $1.4 \times 10^6~\text{yrs}$, in other words,  about $0.6$~Myrs ago. 

\section{Three-Body Dynamical Evolution for pre-main sequence stars}\label{sec:methodology}

Here we discuss the secular three body evolution in the presence of pre-main sequence contraction of the stellar radius and General Relativity (GR) precession and tides. We begin by describing the EKL evolution and our Monte Carlo implementation. In Section \ref{sec:mass} we discuss the possibility of mass transfer during Roche-limit crossing and calculate the resultant temperature of the stars. 

\subsection{Three-Body Dynamical Evolution}\label{sec:dynamics}
We study the hierarchical three-body dynamics, tidal, general relativity, and pre-main-sequence evolution of the binary system Par 1802 in the presence of a tertiary companion. We model the dynamical evolution of these systems by performing Monte-Carlo simulations of $7500$ different initial systems for $2$~Myrs, the approximate age of the Par 1802 system \citep{Gomez+12,Stassun+14}. Our Monte-Carlo implementation first involves randomly selecting $7500$ different sets of initial conditions (that satisfy the stability criteria and constraints described in Section~\ref{sec:ics}). We then solve the three-body secular equations up to the octupole level of approximation \citep[following][]{Naoz+13sec} through numerical integration with the Runge-Kutta 4th order method. We note that a variable step-size was used in our numerical integration due to the presence of highly eccentric orbits. Physical processes such as tidal effects and GR precession for the inner and outer orbits \citep[e.g.,][]{Naoz+13GR} are included in the numerical integration. Additionally, the effect of shrinking stellar radius during the pre-main-sequence evolution of an M-dwarf, which is of particular interest to the case of the Par 1802 binary, is included in the numerical integration.

The quadrupole level of approximation timescale for such a system is
\begin{equation}
t_k\sim \frac{16}{15}\frac{\sqrt{m_1 + m_2}}{\sqrt{G}m_3} \frac{a_2^3}{a_1^{3/2}} \ (1-e_2^2)^{3/2} \ ,
\end{equation}
where $G$ is the gravitational constant, $a_1$ ($a_2$) are the inner (outer) semi-major axis, and $e_1$ ($e_2$) are the inner (outer) eccentricity. 
The tidal evolution of the inner binary is determined following \citet{Eggleton2001} through differential equations that account for distortion due to both tides and rotation. Tidal dissipation follows the equilibrium tidal model in \citet{Egg1998}, with viscous time scale $t_V = 0.5$ for all simulations \citep[see][for full set of equations]{Naoz16}. This is a simplified treatment of tidal effects as $t_V$ is expected to vary with stellar radius and mass. However, the dependence of $t_V$ on radius and mass is unknown. The apsidal motion constants (which is twice the Love parameter) for this polytrope is $0.014$. Hence, this tidal evolution approach is adopted in our integration since it is self consistent with the secular framework, accounts for precession due to tidal torques and oblateness, and enables us to qualitatively understand the physical effect of tides in the inner binary. 

While we adopted equilibrium tides, dynamical tides were shown to have significant effect on the  orbital evolution \citep[see for review][]{Mathis+19}. For example, in a fully convective M-type star such as Par 1802, dynamical tides in the convection zone results in a frequency-dependent tidal dissipation \citep[e.g.,][]{Greenberg+09, Lai+18} that, at resonances, may be significantly larger than equilibrium tide dissipation \citep{Braviner+15}. Dynamical tides have numerous potential consequences on dynamical evolution \citep{Witte+02, Auclair+15}, such as causing the binary orbit to rapidly circularise, abrupt shrinking of $a$, and changes to the orbital period. However, we stress that the equilibrium tide model may capture the qualitative behaviour of the system \citep[e.g.,][]{Lai+18}

\subsection{Initial conditions \label{sec:ics}}
\begin{figure}
	\includegraphics[width=\columnwidth]{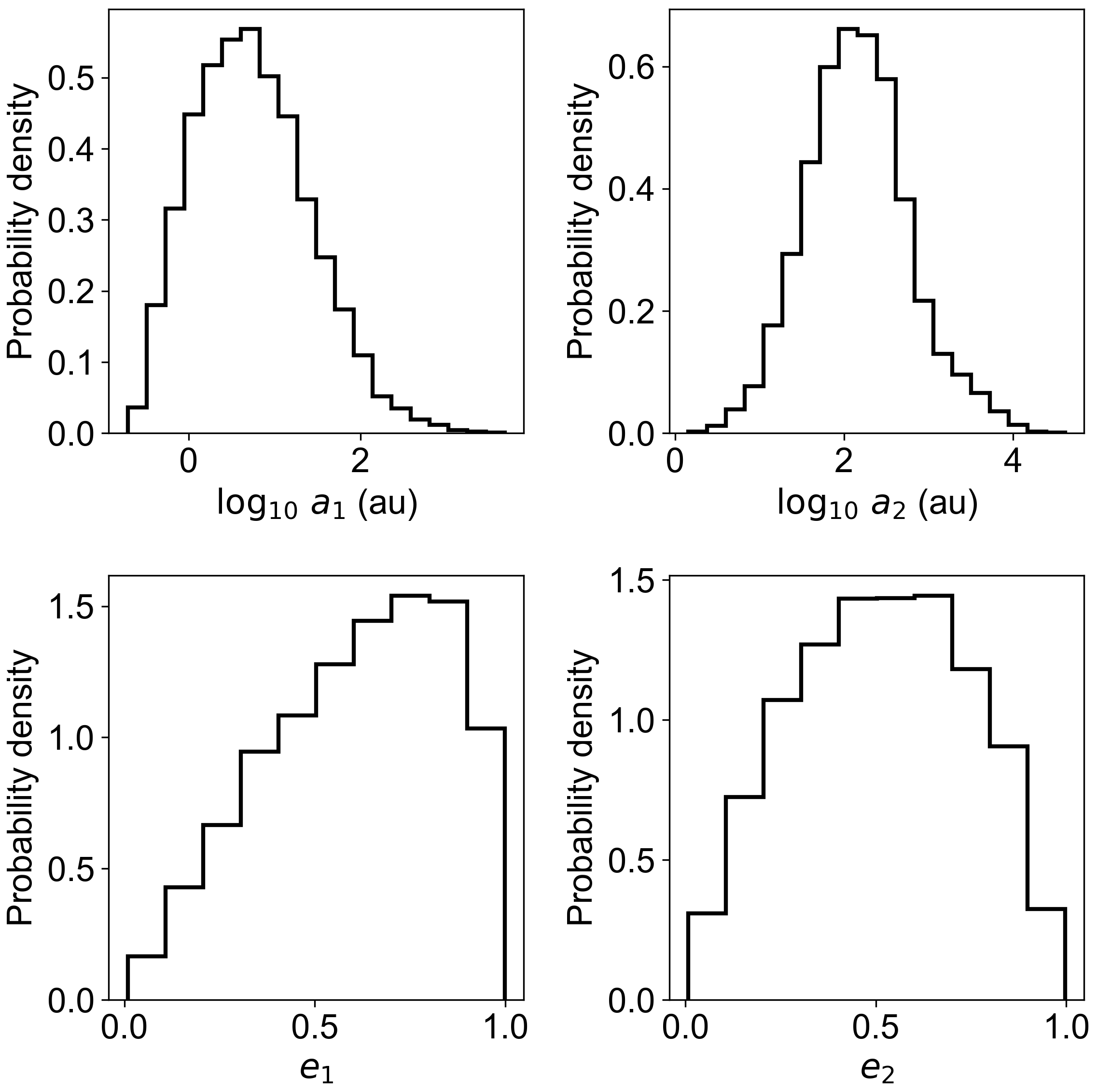}
	\caption{\textbf{Initial conditions for semi-major axes and eccentricities.} The left and right columns show the distribution of $a$ and $e$ of, respectively, the inner and outer binary. Dynamical stability criteria (see text) are accounted for in these distributions.}
	\label{fig:initial}
\end{figure}

The numerical setup incorporates data from observations of Par 1802, selects initial distributions for unknown orbital quantities, and finally applies constraints and stability criteria. 
The masses of the constituent stars of Par 1802 ($m_1$, $m_2$) are taken to be $0.391$~\(M_\odot\) and $0.385$~\(M_\odot\) from \citet{Gomez+12}, and the radii ($R_1$, $R_2$) taken to be $11.71$~\(R_\odot\) and $11.64$~\(R_\odot\), respectively from \citet{Baraffe+98}. The stellar spins for both stars in the inner binary are taken to be $2$ days, and the spin-orbit angle was randomly sampled from an uniform distribution. The mass of the third outer star $m_3$ was randomly sampled from an uniform distribution between $0.1$~\(M_\odot\) and $0.8$~\(M_\odot\). 

The eccentricity of the inner (outer) orbit $e_1$ ($e_2$) were sampled from a thermal distribution between $0$ and $1$. The inclination $i$ between the inner and outer orbit's angular momenta was assumed to be isotropic (i.e., uniform in cosine). The period of the inner and outer orbit was assumed to be the log-normal distribution of \citet{Duquennoy+91}. Note that this period distribution represents the final periods of binaries population, rather than the initial one. However, if we extrapolate for a recent study for more massive stars, the final period distribution, at wide orbits, is a signature of the birth distribution \citep{Rose+19}.

\begin{figure*}
	\includegraphics[width=\linewidth]{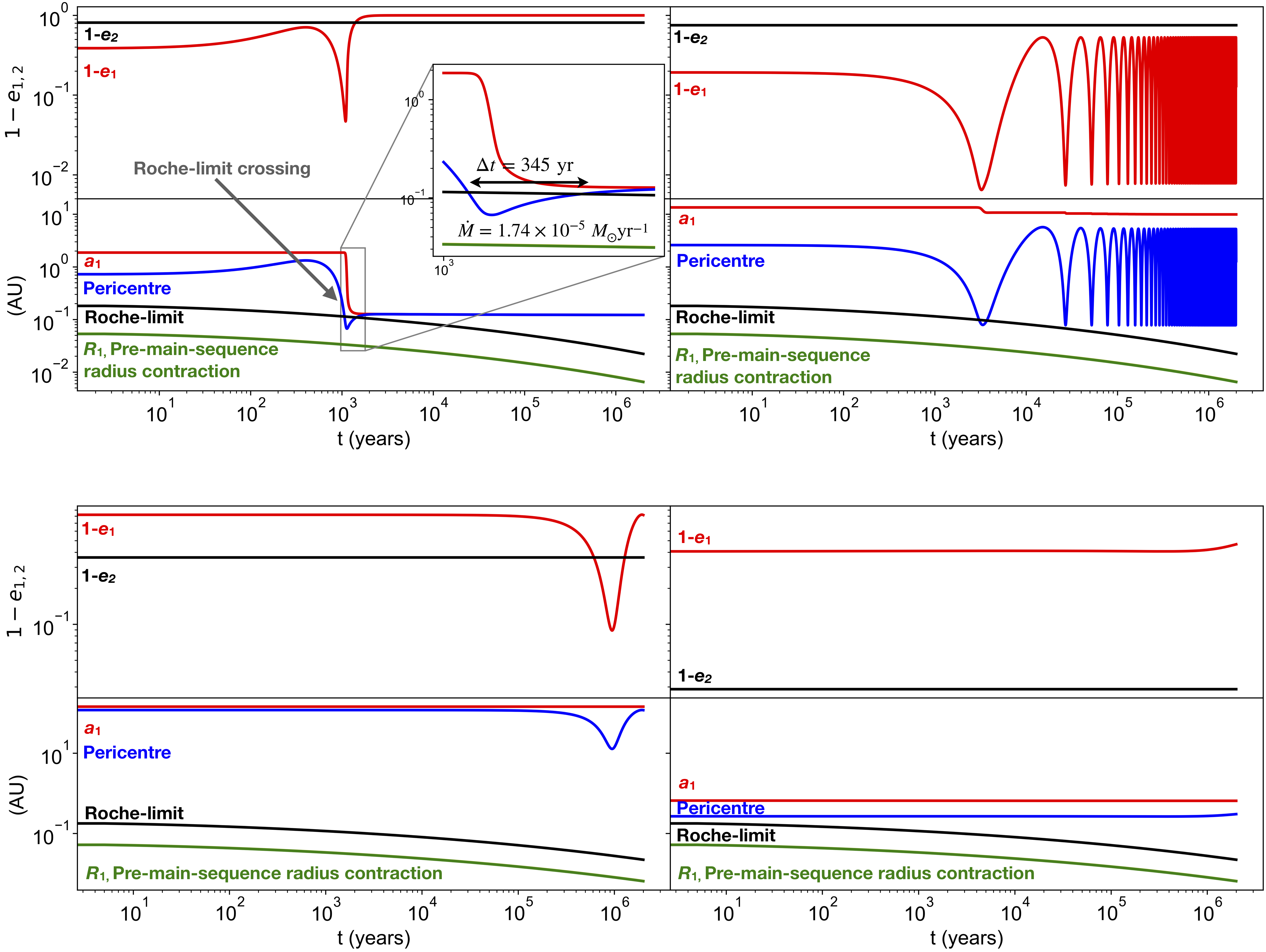}
	\caption{\textbf{Archetypal systems.} We show the time evolution of 4 triple systems. In the top row, we show two Roche-limit crossing systems: with a single eccentricity excitation (left), and with long term eccentricity oscillations (right). In the bottom row, we show two Non-Roche-limit crossing systems. For each of the 4 systems, the top panel shows $1-e$ of the inner and outer binary and the bottom panel shows the inner semi-major axis, pericentre, the Roche-limit, and $R_1$, the radius of the more massive inner binary star. When plotted, $R_2$ is visually indistinguishable from $R_1$ and is therefore omitted from the figure. All systems have $m_1 = 0.391~M_\odot$, $m_2 = 0.385~M_\odot$, and spins of $2$ days. The initial conditions are (from left to right on top row, then left to right on bottom row): $m_3 = 0.25, 0.36,0.67,0.49~M_\odot$, $a_1 = 1.87, 13.4, 144, 0.658$~au, $a_2 = 14.7,108,2220,1195$~au, $e_1 = 0.62,0.81,0.18,0.59$, $e_2 = 0.19,0.25,0.64,0.97$, $g_1 = 131.6, 231.8,263.5,128.1^{\circ}$, $g_2 = 136.0,123.9,255.3,37.29^{\circ}$, and $i = 86.4,101,112,121 ^{\circ}$.}
	\label{fig:archetype}
\end{figure*}

We also adopt the following stability criteria. The first represents the 
relative importance of the octupole and quadrupole terms in the system's Hamiltonian \citep{Naoz16}:
\begin{equation}
\epsilon = \frac{a_1}{a_2} \ \frac{e_2}{1-e_2^2} < 0.1 \ .
\label{equ:epsilon}
\end{equation}
The second stability criteria follows \citet{Mardling+01}:
\begin{equation}
\frac{a_2}{a_1} > 2.8 \left(1+\frac{m_2}{m_1 + m_2}\right)^{2/5} \frac{(1+e_2)^{2/5}}{(1-e_2)^{6/5}} \left(1-\frac{0.3 i}{180^{\circ}}\right) \ .
\label{eq: marv}
\end{equation}
Equation~\ref{equ:epsilon} is numerically similar to this criteria \citep{Naoz+13sec}. It is noted that the dynamics of systems in the upper limit of Equation~\ref{equ:epsilon} may be dominated by effects other than the EKL mechanism, especially for a strong perturber \citep{Antonini+14, BodeWegg+14}. Our work does not take these cases into account because Equation~\ref{equ:epsilon} and~\ref{eq: marv} are numerically very close, and therefore it is expected that the vast majority of stable systems can be faithfully described by the EKL mechanism. We also note that these stability criteria deem a system unstable if at any point in time it violates the stability criterion, but do not provide a timescale for growth of the instability \citep[e.g.,][]{Myllari+18}. In practice we expect more systems to be stable for $2$~Myr of our integration timescale than considered here. 

Additionally, we require the pericentre distance to be greater than initial Roche-limit, which ensures that the inner binary does not merge before the tertiary companion can affect the system:
\begin{equation}
a_1(1-e_1) > R_{\rm Roche, 1} \ .
\end{equation}
Together, these constraints and stability criteria allow us to model systems in which the EKL mechanism may facilitate tidal heating during the pre-main-sequence evolution of the system. We depict the initial conditions of the system in Figure \ref{fig:initial}, after these criteria have been applied. 

\section{Three-body evolution Results}\label{sec:results}
\subsection{\label{sec:dynam_outcomes}Inner Orbit Dynamical Outcomes}
We identified two general dynamical outcomes. As expected, the outcome is dependent on the presence and strength of eccentricity excitations. EKL in the presence of tides tends to circularise and tighten systems. Larger eccentricity excitations, due to the EKL mechanism, results in a more efficient production of such systems. The different dynamical outcomes are depicted in Figure~\ref{fig:archetype}, and are summarised in Figure \ref{fig:inner}. In the latter Figure we consider the initial distribution of $a_1$ (top) and $e_1$ (bottom) and over-plot the $15\%$ of systems that crossed their Roche-limit at any point throughout the $2$~Myr evolution (blue line). The green vertical lines in Figure  \ref{fig:inner} represents the observed parameters for Par 1802. We then use the observed $a_1$, up to a factor of $2$, to constrain the orbital parameters of the third companion.

\subsubsection{Roche-limit crossing systems}
\label{sec:roche}
About $15\%$ of all systems crossed their Roche-limit.  
We identify two subtypes of Roche-limit crossing systems:
\begin{itemize}
    \item \textbf{Single eccentricity excitation.} Depicted in top left panel in Figure~\ref{fig:archetype}. These systems are in the most efficient part of the parameter space of EKL. Here, the eccentricity undergos a single dip throughout the evolution, with pericentre dipping below the Roche-limit of the inner binary for a time $\Delta{t}$ and then exiting the Roche-limit. Mass transfer may occur within the Roche-limit (see Section \ref{sec:mass}).
    These systems are of interest since Par 1802 is observed to be circularised ($e_1 = 0.0166$) with small semi-major axis ($a_1 = 0.0496$~au) \citep{Gomez+12}.
    \item \textbf{Long term eccentricity oscillations.} Depicted in top right panel in Figure~\ref{fig:archetype}. In these systems EKL oscillations take place, but the eccentricity excitation is not high enough to cause the system to quickly circularise. Tides may cause an increase of $a_1$. In general, for these systems the inner binary does not circularise and tighten, and therefore these systems are not of interest as they do not match observations. 
\end{itemize}

\subsubsection{Non-Roche-limit crossing systems}
In many cases, the EKL eccentricity excitation are either taking place on longer timescales than the age of the system (e.g., bottom right panel in Figure ~\ref{fig:archetype}), or had not yet reached high eccentricity values (bottom left panel in Figure~\ref{fig:archetype}). These systems are not of interest since the inner binary does not circularise or tighten, and its final configuration does not match observations.

\begin{figure}
	\includegraphics[width=\columnwidth]{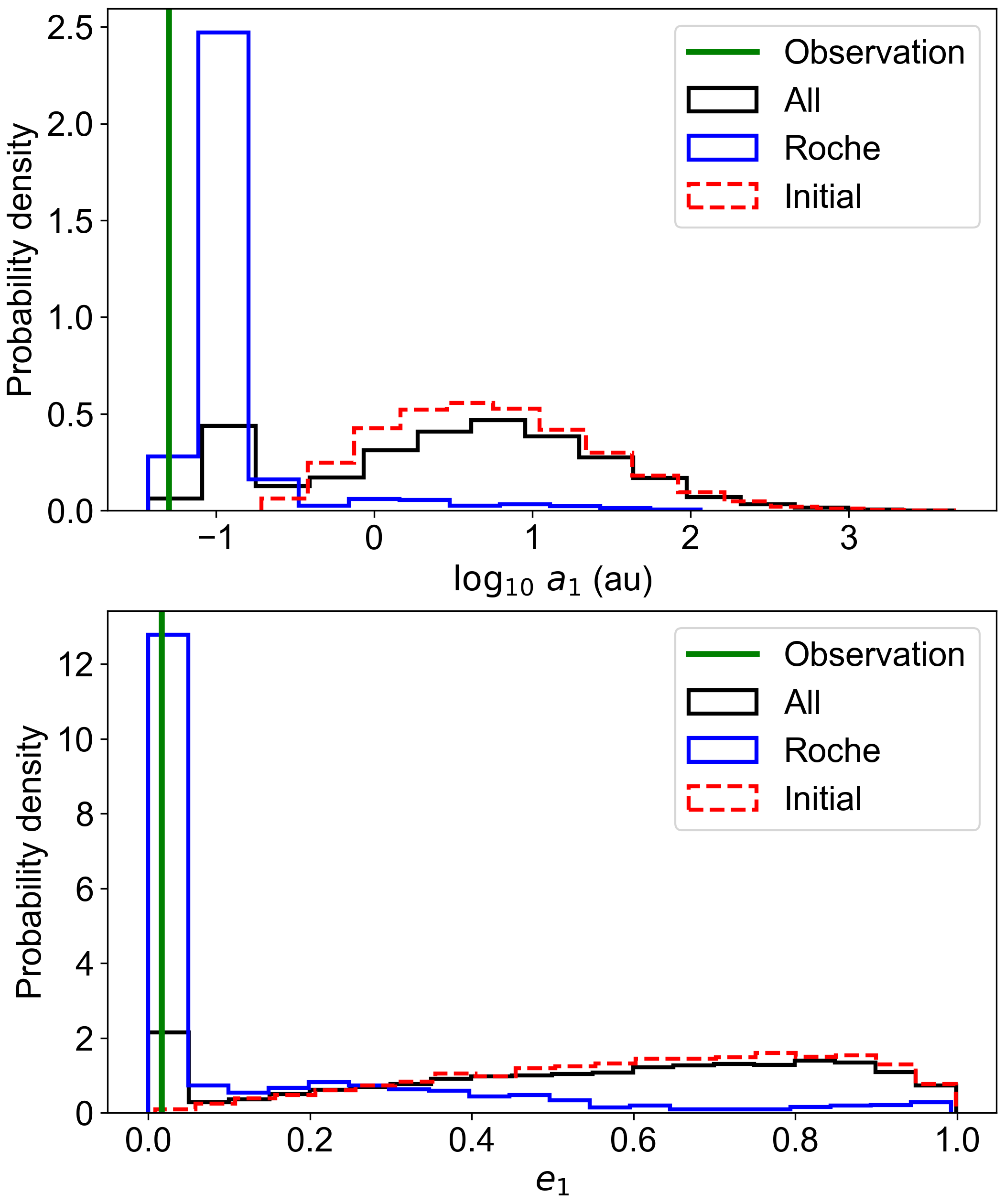}
	\caption{\textbf{Inner binary orbit parameters.} The top and bottom panels show the distribution for, respectively, the semi-major axis and eccentricity of the inner binary. Black lines shows the final distribution for all systems, and blue lines show the final distribution for systems that crossed the Roche-limit. Red dashed lines show the initial distribution for all systems, and the observed quantities \citep{Gomez+12} are in green.}
	\label{fig:inner}
\end{figure}
\begin{figure}
	\includegraphics[width=\columnwidth]{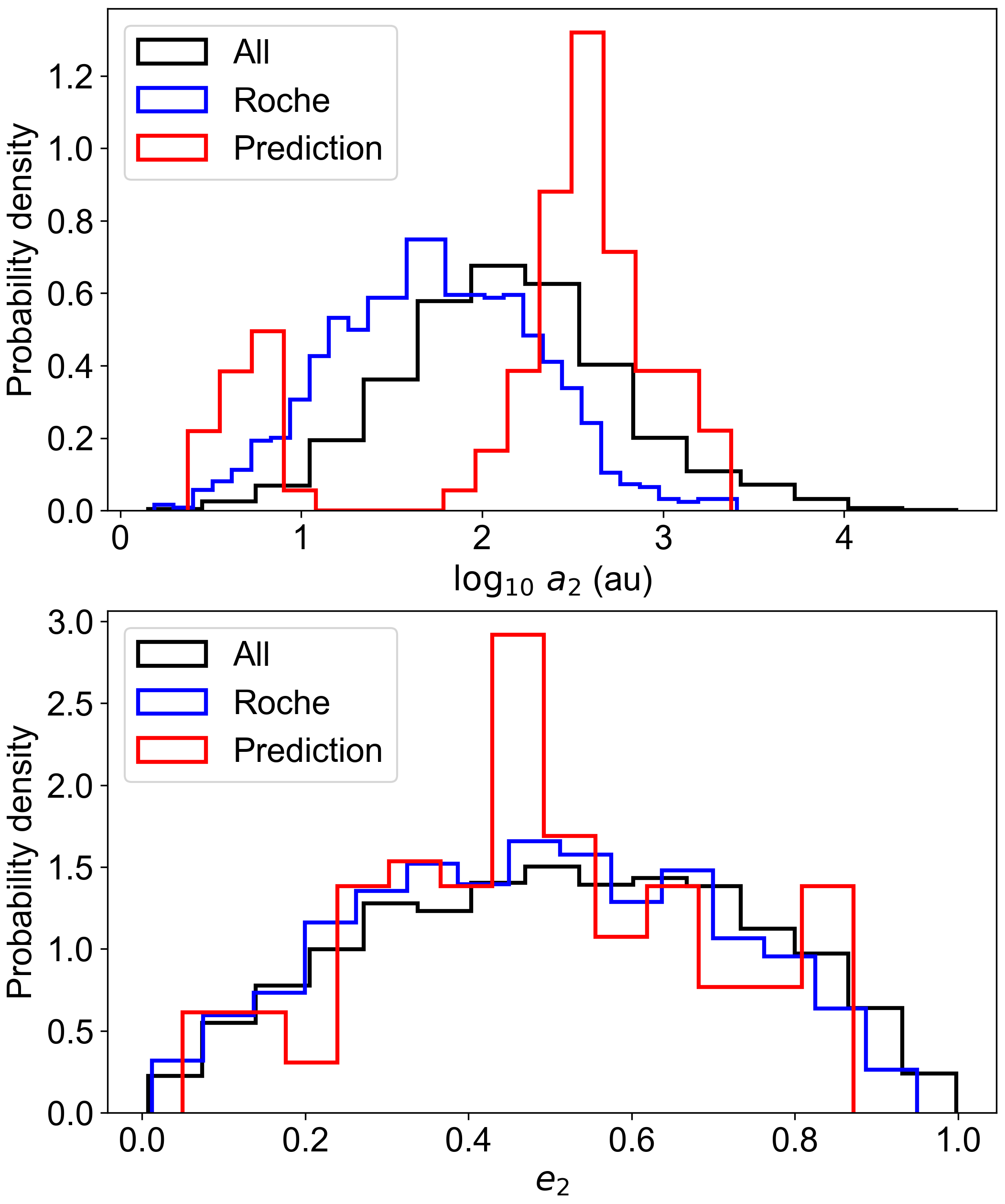}
	\caption{\textbf{Outer binary orbit parameters.} The top and bottom panels show the distribution for, respectively, the semi-major axis and eccentricity of the outer binary. Black lines shows all systems, and blue lines show only systems that crossed the Roche-limit. Red lines show Roche-limit crossing systems constrained by the observed value for $a_1$. The cluster of constrained systems with $a_2 < 30$~au correspond to $e_2 < 0.5$, and represents systems with a moderately tightened and circularised outer orbit.}
	\label{fig:outer}
\end{figure}

\subsection{Tertiary predictions}
We focus on Roche-limit crossing systems since their behavior matches the observations of \citet{Gomez+12} in which the inner binary is circularised and tightened. We show in Figure~\ref{fig:inner} that the vast majority of Roche-limit crossing systems (those with a single eccentricity excitation) have a final inner binary configuration that is circularised and tightened. Qualitatively, we note, that these  systems should be a generic result of the EKL evolution. About $10\%$ of all systems that cross their Roche-limit are within in a factor of $2$ of Par 1802 observed semi-major axis of $0.0496 \pm 0.0008$~au and have a circularised orbit. We use this subset to constrain the parameter space for the Par 1802 hypothetical companion. We choose to constrain our predictions using a factor of $2$ due to the uncertainties of our model, such as in the prefactor of the Roche-limit in Equation~\ref{eq:Roche} and the choice of our viscous timescale in Section~\ref{sec:dynamics}.  

We predict the median semi-major axis for Par 1802 outer binary to likely be $350$~au. 

As shown in the top panel of Figure~\ref{fig:outer}, Roche-limit crossing systems (blue) have $a_2$ that ranges between $2-2500$~au with a median of $55$~au. This is distinctly different from the distribution for $a_2$ for all simulated systems (black), which has a median of $140$~au. Due to the young age of Par 1802 (and thus the short time of simulation), the final distribution for $a_2$ for all systems (black) closely resembles the initial conditions (e.g., top right panel in Figure~\ref{fig:initial}). This result is consistent with the result for massive stars found by \citet{Rose+19}.

The Par 1802-like systems (the $10\%$ of systems within factor of $2$ of the observed $a_1$) have a bi-modal distribution in $a_2$. The low value $a_2$ peak (median of $\sim 5$~au) corresponds to low $e_2$ systems. The higher $a_2$ values represent the majority of systems and has a median of $\sim 375$~au.

As expected the final distribution of $e_2$ resembles the initial distribution. The eccentricity of Par 1802-like system's outer orbit corresponds to the outer orbit semi-major axis, with the rule of thumb that lower eccentricity requires lower (closer in) semi-major axis values. The median eccentricity in this case is $0.5$ (bottom panel in Figure \ref{fig:outer}). 

\begin{figure}
	\includegraphics[width=\columnwidth]{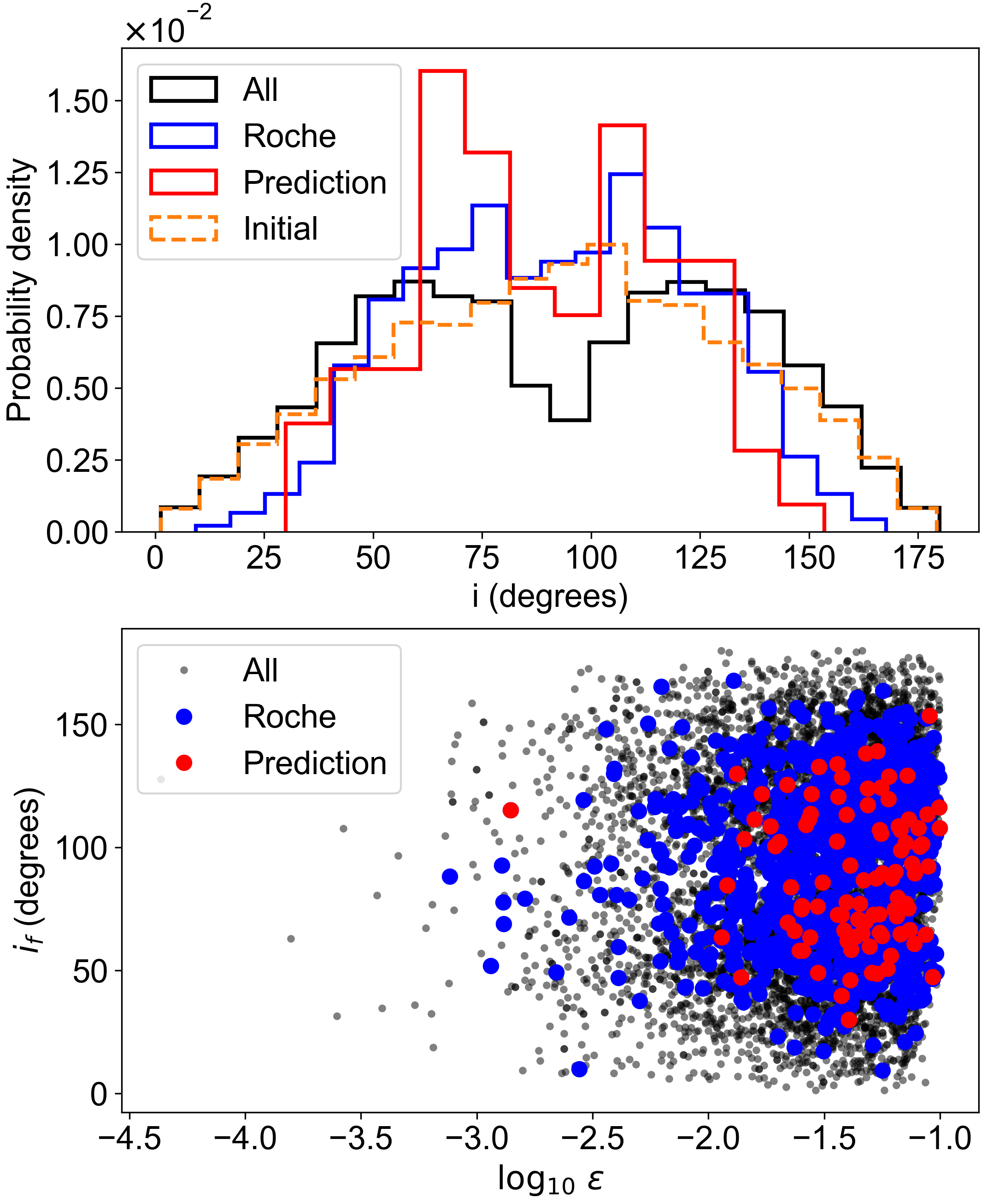}
	\caption{\textbf{Inclination between inner and outer binary angular momenta, and $\epsilon$.} The top panel shows the distribution of the inclination, with blue marking all Roche-limit crossing systems, dashed orange for the initial distribution for all systems, and red marking Roche-limit crossing systems constrained by the observed value for $a_1$. The bottom panel shows the final inclination for all systems $i_f$ as a function of $\epsilon$, with all Roche-limit crossing systems marked in blue and Roche-limit crossing systems constrained by the observed value for $a_1$ marked in red.}
	\label{fig:incl}
\end{figure}

The inclination of the system (top panel in Figure \ref{fig:incl}) also behaves as expected, and we find a bi-model distribution \citep[consistent with][]{Fabrycky+07,NF}. However, due to the short timescales of the integration, the bi-modal distribution is more closely located near $90^\circ$ rather than near the Kozai-angles, being centered at about $70^{\circ}$ degrees and $110^{\circ}$ degrees. The inclination of the system is correlated with $\epsilon$ (bottom panel of Figure \ref{fig:incl}), a parameter that describes the strength of the EKL mechanism. As expected, the EKL mechanism is efficient for systems that cross the Roche-limit and particularly efficient for Roche-limit-crossing systems that are within a factor of 2 of the observed $a_1$.

\section{Interaction at Roche-Limit}\label{sec:interaction}
\subsection{\label{sec:mass}Mass transfer}
At the Roche-limit, mass transfer occurs. As a proof-of-concept we assume a conservative mass inversion where the binary mass
\begin{equation}
    m_1 + m_2 = 0.776~M_\odot = \text{constant} \ ,
\end{equation}
and that mass transfer only occurs while the inner binary stars are within the Roche-limit. We note that \citet{Eggleton1983} adopted the Roche Lobe radius (where a surface mass element of a star becomes unbound) as the criterion for mass transfer, with mass transfer occurring slightly earlier. Here, we instead adopt the Roche-limit as the criterion for mass transfer, which is more conservative (smaller $\Delta{t}$) than the Roche Lobe radius criterion by a small factor of $\sim 1.2$. This factor is well within the uncertainties in the model, and therefore will not lead to any significant qualitative changes to the result. 

Thus, the mass transfer rate is loosely defined as:
\begin{equation}
    \dot{M} = \frac{\Delta{M}}{\Delta{t}} \ ,
    \label{eq:mdot}
\end{equation}
where $\Delta M=m_1-m_2$. 
In our proof-of-concept model, the mass transfer rate varies for different lengths of time $\Delta{t}$ spent within the Roche-limit. Since we assume mass inversion, the fraction of binary mass transferred is constant between different systems. Realistically we may expect a constant mass transfer rate rather than constant fraction of binary mass transferred.

\begin{figure}
	\includegraphics[width=\columnwidth]{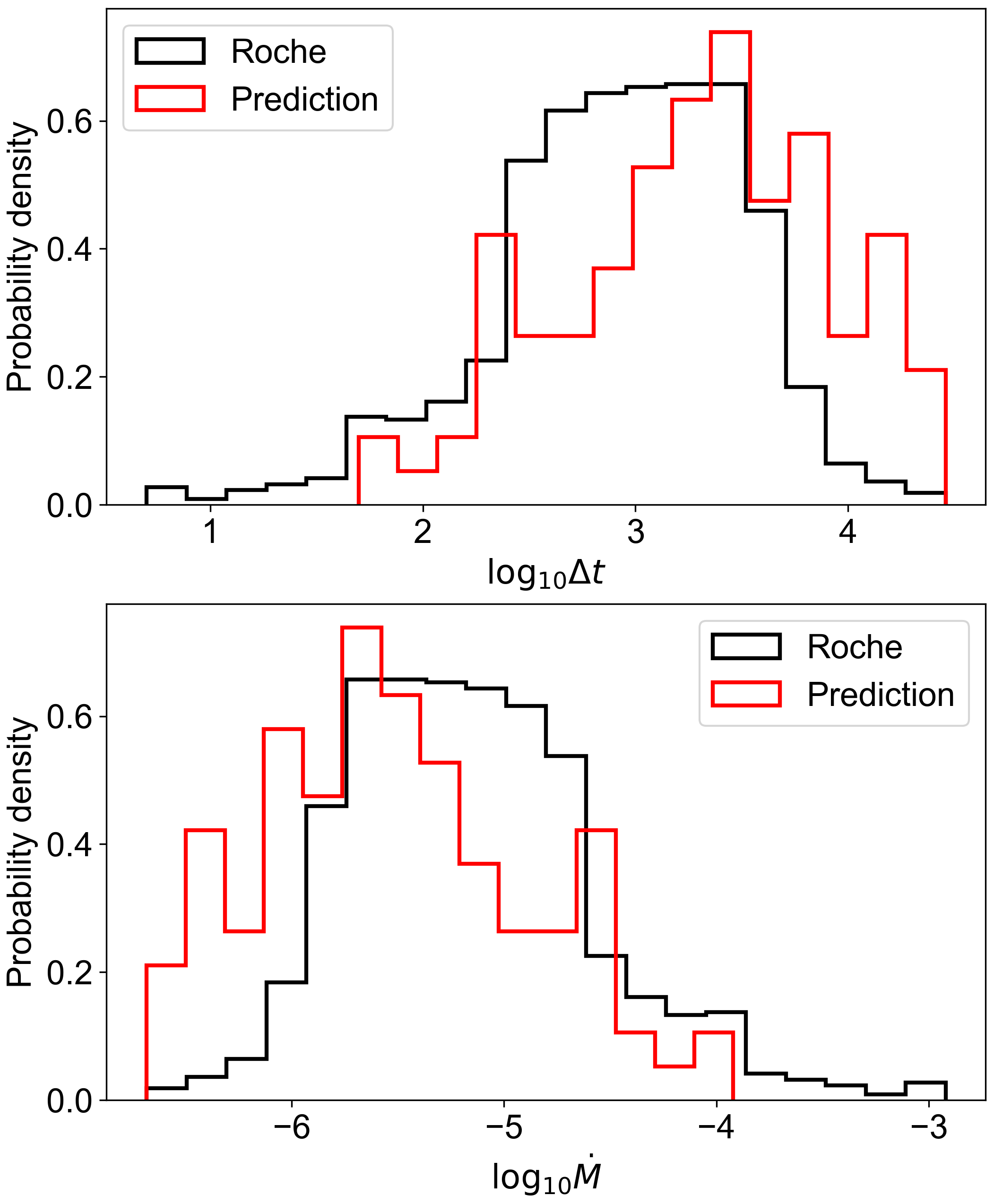}
	\caption{\textbf{Mass transfer characteristics for Roche-limit crossing systems.} The top panel shows the amount of time $\Delta{t}$ Roche-limit crossing systems spent within the Roche-limit, and the bottom panel shows the mass transfer rate for all Roche-limit crossing systems determined using Equation~\ref{eq:mdot}. Roche-limit crossing systems constrained by the observed value for $a_1$ are marked in red.}
	\label{fig:roche}
\end{figure}

The time that the systems spend in each-other Roche-limit is about $1000$~yr which results in a constant mass transfer rate of 
$\sim 6\times 10^{-6}$~M$_\odot$~yr$^{-1}$ (see Figure \ref{fig:roche}). The Par 1802-like systems spend about a factor of $3$ longer time inside the Roche-limit, which results in a mass transfer rate of $\sim 3\times 10^{-6}$~M$_\odot$~yr$^{-1}$. These mass transfer rates are not unreasonable to assume and we thus proceed with this proof-of-concept calculation. We note that alternative models exist \citep{Hamers+19,Dosopoulou+16,Sepinsky+07} that incorporate orbital evolution, and will be included in future work.

\subsection{\label{sec:heat}Tidal Heating}
As the pre-main sequence stars undergo mass transfer they also feel the tidal forces that shrink and circularised the orbit (see for example Figure \ref{fig:archetype}).  
The tidal model adopted throughout the three body dynamical evolution followed the equilibrium tidal model from \citet{Hut80} that assumes a constant viscous time \citep[e.g.,][see the latter for a complete set of equations]{Egg1998,Eggleton2001,Naoz16}. However we are particularly interested in the response of the pre-main sequence stars to the energy deposited within them due to tides during the Roche-limit crossing period.

Tides on these polytropes that take orbital energy are transferred into heat in the star. This may cause the inflation of the star. Here we adopt a constant phase-lag tidal heating model \citep{Goldreich+66,Wisdom+08,Ferraz+08}, often used for brown dwarfs \citep[e.g.,][]{Heller+10}. Since the stars in questions did not begin their thermonuclear fusion reaction, this tidal model is adequate.\footnote{Note that dynamical tides have a significant effect on tidal dissipation, and may allow more efficient tidal heating \citep[see for review][]{Mathis+19}. The tidal heating model used here is based on the static equilibrium tidal model and is sufficient for our proof-of-concept analysis.} We note that the constant phase-lag model adopts a constant quality factor for each of the stars $Q_j$, with $j=1,2$. 
However, since $a_1$ and $R_j$ of the stars ($j=1,2$) vary in time, the quality factor at the onset of Roche-lobe crossing is dramatically different than the quality factor at the existing point. The evolution of $Q$ strongly impacts the tidally driven orbital evolution of systems \citep{Heller+18}.

For the following proof-of-concept calculation we adopt two quality factors to calculate the excess temperature due to tidal effects. First, we consider the relation between the quality factor and the viscous time: 
\begin{equation}
Q_j = \frac{4}{3}\frac{k_1}{(1+2k_1)^2}\frac{Gm_j}{R_j^3}\frac{t_V}{n_1} \ ,
\label{eqn:Q}
\end{equation}
from \citet{Naoz+16}, with constant viscous timescale $(t_V)_j=0.5$ typical for M-dwarfs, classical apsidal motion constant $k_1 = 0.014$, and where the mean motion is $ n_1 = {2\pi}/{P_1}$. Since $R_j$ and $a_i$ varies in time, $Q_j$ varies in time.

Tidal heating takes place throughout the tidal evolution. However, it is most significant when the stars are within each-other Roche-Lobe regimes (see upper panel of Figure~\ref{fig:roche} for time scale). We thus limit ourselves to two relevant points in time of the evolution: the onset of the Roche-limit, and exit from this regime. We find that for the majority of cases $Q_{j,\rm enter}>>Q_{j,\rm exit}$, where the subscript ``enter'' (``exit'') refers to entering (exiting) the Roche-limit. While tidal heating takes place for both stars, the $3\%$ mass difference results in slight changes in the amount of heating per unit mass, with the less massive star being heated more.

Under the tidal heating, constant phase-lag model, the potential of the less massive body is expressed in terms of periodic contributions of tidal frequencies at various phase-lags \citep[e.g.,][]{Heller+10}, which can be expanded in terms of these phase lags. The phase lags $\epsilon_{\eta,j}$ for the $j$th body with $\eta=0,1,2,5,8,9$ are:
\begin{align}
Q_j \ \epsilon_{0,j} &= \Sigma(2\Omega_{\text{s},j} - 2n_1) \nonumber\\
Q_j \ \epsilon_{1,j} &= \Sigma(2\Omega_{\text{s},j} - 3n_1)\nonumber\\
Q_j \ \epsilon_{2,j} &= \Sigma(2\Omega_{\text{s},j} - n_1)\nonumber\\
Q_j \ \epsilon_{5,j} &= \Sigma(n_1)\nonumber\\
Q_j \ \epsilon_{8,j} &= \Sigma(\Omega_{\text{s},j} - 2n_1)\nonumber\\
Q_j \ \epsilon_{9,j} &= \Sigma(\Omega_{\text{s},j}) \ ,
\end{align}
with $\Sigma(x)$ defined as the algebraic sign of $x$:
\begin{equation}
    \Sigma(x) = +1 \lor -1 \ ,
\end{equation}
and $\Omega_{\text{s},j}$ is the rotational period associated with the spin of the stars.

The change in orbital energy of star $1$ due to star $2$ (or vice versa) is then: % \citep{Heller+10}
\begin{eqnarray}
    \dot{E}_{\text{orbit}, j} &=& \frac{3k_LGm_l^2R_j^5}{8a_1^6}n(4\epsilon_{0,j} + e_1^2[-20\epsilon_{0,j}+\frac{147}{2}\epsilon_{1,j}+\frac{1}{2}\epsilon_{2,j} \nonumber \\
    &-& 3\epsilon_{5,j}]-4\sin^2{\psi_j}[\epsilon_{0,j}-\epsilon_{8,j}]) \ ,
\end{eqnarray}
where $j,l=1,2$, and the change in rotational energy is
\begin{eqnarray}
    \dot{E}_{\text{rotation}, j}& =&  -\frac{3k_LGm_l^2R_j^5}{8a_1^6}\Omega_{\text{s},j}(4\epsilon_{0,j} + e_1^2[-20\epsilon_{0,j}+49\epsilon_{1,j}+\epsilon_{2,j}]\nonumber \\
   & +&2\sin^2{\psi_j}[-2\epsilon_{0,j}+\epsilon_{8,j}+\epsilon_{9,j}]) \ ,
\end{eqnarray}
where $k_L = 0.5$ is the dynamical Love parameter and $\psi_j$ is the obliquity of the $j$th body. 
All of the relevant variables are obtained through the dynamical evolution results described in Section~\ref{sec:dynam_outcomes}, and the masses of the stars invert between Roche-limit entry and exit (see Section~\ref{sec:mass}). Finally the change in total energy due to tides is
\begin{equation}
    \dot{E}_j = -(\dot{E}_{\text{orbit}, j} + \dot{E}_{\text{rotation}, j}) \ . \label{eq:lum}
\end{equation}

The temperature change for each body due to tidal heating is then evaluated from the Stefan-Boltzmann equation
\begin{equation}
    dT_j = \left(\frac{\dot{E}_j/2}{4\pi{}R_j^2\sigma_{\text{SB}}}+T_{\text{eff},j}^4 \right)^{1/4} - T_{\text{eff},j} \ ,
\end{equation}
where $\sigma_{\text{SB}}$ is the Stefan-Boltzmann constant, and $T_{\text{eff},j}$ is the temperature at Roche-crossing for the $j$th body of mass $m_j$ adopting \citet{Baraffe+98} model.
We then estimate the above equation at two times: when the system enters and exits the Roche-Lobe regime\footnote{In reality, heating takes place throughout the time the stars spend in the Roche-limit. However, since we find that the maximum heating takes place when the stars exit their Roche-limit, we evaluate heating only at these two points in time.}. In other words, $dT_j=dT_{j,\rm {enter}} + dT_{j,\rm exit}$, where subscript enter (exit) refers to entering (exiting) the Roche-Lobe regime. We note that $T_{j,\rm eff}$ at the entering and exiting points of the Roche-limit are roughly the same because the short amount of time the stars spends in each-other Roche-lobe regime (see Figure \ref{fig:roche}).
The temperature difference is then estimated by:
 \begin{equation}
     \Delta{T} = (dT_1 + T_{\text{eff},1}) - (dT_2 + T_{\text{eff},2}) \ .
 \label{eqn:temp}
 \end{equation}

\begin{figure}
	\includegraphics[width=\columnwidth]{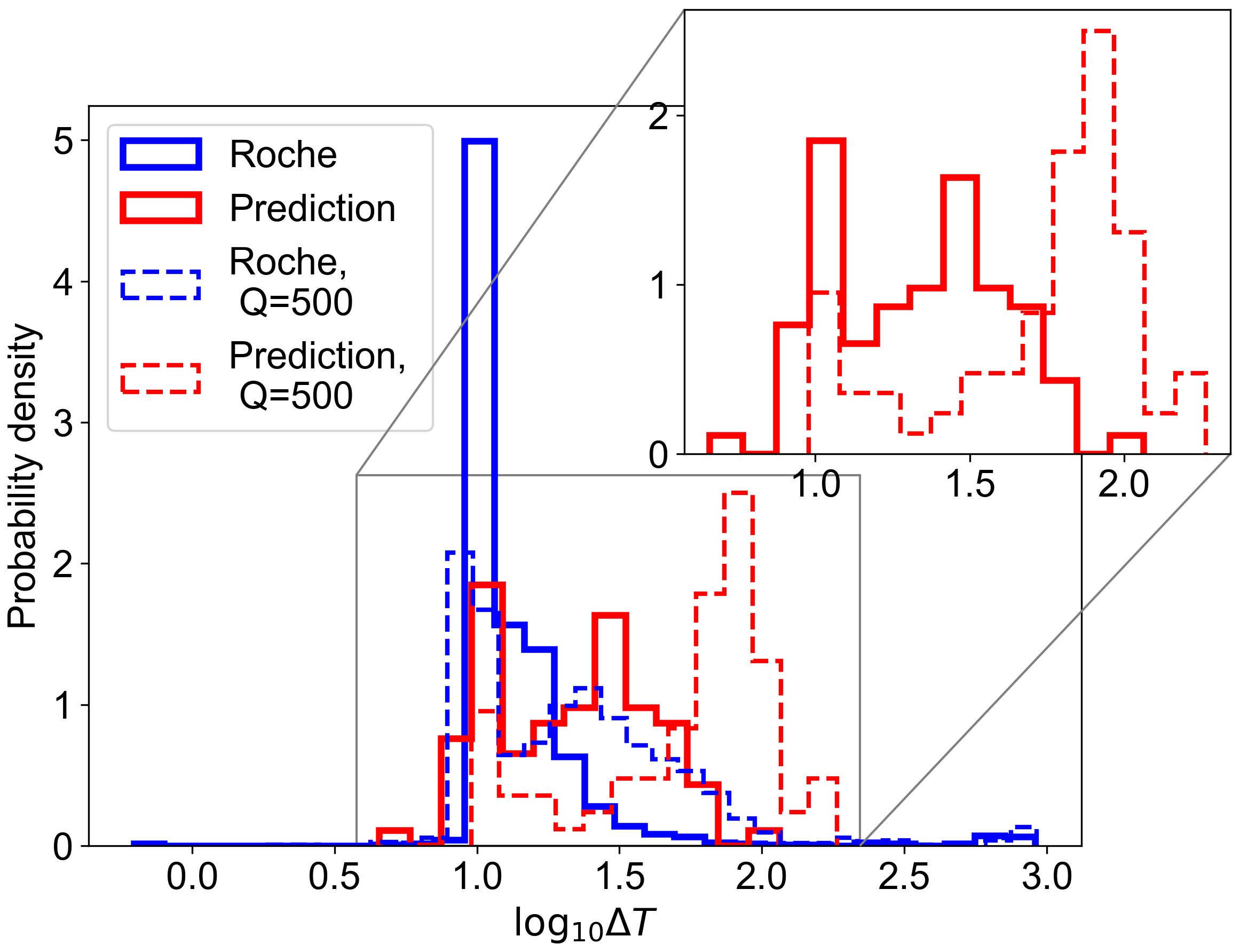}
	\caption{\textbf{Temperature difference between inner binary stars.} The solid lines show the temperature difference between the inner binary stars determined using Equation~\ref{eqn:temp} through the method detailed in Section~\ref{sec:heat}. The quality factor adopted in this calculation follows Equation~\ref{eqn:Q}, and varies with different $a_1$ and $R_j$. Roche-limit crossing systems are in blue, and Roche-limit crossing systems constrained by the observed value of $a_1$ are in red. The maximum $\Delta{T}$ for all Roche-limit crossing systems is $910$~K, and the maximum predicted $\Delta{T}$ for Roche-limit crossing systems constrained by the observed value of $a_1$ is $115$~K. Additionally, the temperature difference for an assumed constant quality factor $Q=500$ is overplotted in dashed lines. With this constant $Q$, the maximum $\Delta{T}$ for all Roche-limit crossing systems is $920$~K. After constraining with the observed value of $a_1$, the maximum $\Delta{T}$ is $180$~K.}
	\label{fig:temp}
\end{figure}

In Figure \ref{fig:temp} we show a histogram of the final excess temperatures as a result from the heuristic calculation of tidal heating. We find that Roche-limit crossing systems have a wide range of $\Delta{T}$, with a maximum $\Delta{T}_{\text{max}} = 910$~K (Figure~\ref{fig:temp}). We predict a maximum $\Delta{T}_{\text{predict}} \approx 115$~K.

The luminosity of the two stars is estimated by summing the total energy change due to tides for each star using Equation \ref{eq:lum} at entry and exit from the Roche-limit. The luminosity difference between Par 1802-like stars (with $a_1$ within a factor of 2 from observation) has a median of $\sim 3 \times 10^{25}$~W, which is comparable to the observed luminosity difference of $\sim 7 \times 10^{25}$~W \citep{Gomez+12}.

Since the constant phase-lag tidal heating model typically adopts a constant $Q$, we complete a second set of calculations of $\Delta{T}$ for an assumed constant $Q=500$. We predict a maximum $\Delta{T} \approx 180$~K for constant $Q=500$. Both predictions ($\sim 115$~K for $Q$ following Equation~\ref{eqn:Q}, and $\sim 180$~K for constant $Q=500$) are comparable to the observation of $\Delta{T}_{\text{observed}} \approx 300$~K \citep{Stassun+14}.
With $Q = 500$, the luminosity difference between Par 1802-like stars has a median of $\sim 5 \times 10^{26}$~W, comparable to observation.

\section{Discussion}\label{sec:discussion}
The Par 1802 M-dwarf binary system is an interesting quandary for stellar evolution since one of the component stars has unexpectedly higher temperature compared to the other despite their almost identical (within $3\%$) masses. Previous work by \citet{Stassun+14} and \citet{Gomez+12} have proposed that the temperature difference was caused by the presence of a third star. Here we focused on the dynamical evolution of such a system and have shown that observed quantities can be naturally reproduced. Furthermore, we predict the orbital configuration of such a tertiary companion.

We studied the secular evolution of the binary M-dwarf system Par 1802 in the presence of a third, far away star.  We numerically solved the three-body hierarchical secular equation to the  octupole level of approximation \citep[i.e., the EKL mechanism][]{Naoz16}. Tidal effects, GR precession, and pre-main-sequence evolution were self-consistently included in our simulations. 

During the evolution, the EKL mechanism could cause large eccentricity excitations in the inner orbit, which prompted the inner binary to tighten and circularise (see top left plot in Figure~\ref{fig:archetype}). In some of these  systems, eccentricity excitations caused the inner binary stars to cross their Roche-limits (see Figure~\ref{fig:archetype}). During that time, mass transfer could take place, resulting in the aforementioned temperature difference. As the stars' radii contracted their Roche-radii shrunk, effectively halting mass transfer, and the stars exit the Roche-lobe regime.

We found that $\sim15\%$ of all simulations were Roche-limit crossing systems with the majority ($\sim80\%$) of these having circularised and their separation shrunk due to tides (see Figure~\ref{fig:inner}).
We focused on systems that evolved to within a factor of $2$ of the observed Par 1802 separation (of $\sim 0.05$~au). These systems represented about $10\%$ of all Roche-limit crossing systems, and about $1.5\%$ of all simulated systems. Since our initial conditions for semi-major axis and eccentricity featured broad distributions (with $a_1$ varying between $\sim 0.2-380$~au and $e_1$ varying between $\sim 0.1-0.9$), this $1.5\%$ fraction is statistically significant and suggests that systems similar to Par 1802 can exist in the presence of a tertiary companion.

For these systems similar to Par 1802, we constrain the orbital configuration for the Par 1802 tertiary star. In particular, we predict that the outer orbit is most likely to have a separation of $350$~au (top panel Figure~\ref{fig:outer}). We find that the predicted median eccentricity of the tertiary is $e_2 = 0.5$ (bottom panel Figure~\ref{fig:outer}). The inclination, \citep[as expected, e.g.,][]{NF}, has a bi-modal distribution with peaks centered around $70^{\circ}$ and $110^{\circ}$.

During the time the system crossed its Roche-limit, mass transfer took place and tidal heating was probably most efficient. The two stars of Par 1802 have an observed temperature difference of $\sim300$~K \citep{Gomez+12}. We conducted a proof-of-concept calculation, assuming conservative mass transfer during the Roche-limit crossing time. For simplicity the two inner masses were set to be the observed masses, and thus, to match observations the mass ratio inverted over the course of the evolution. These assumptions determined the mass transfer rate (see Figure~\ref{fig:roche}). Following the \citet{Heller+10} tidal heating model we find about $0.2\%$ of all systems have temperature difference within a factor of $2$ of the observed value. These systems represent a small fraction ($5\%$) of the Par 1802-like systems (i.e., systems with final inner binary separation within a factor of 2 of the observed separation). Since our mass exchange and tidal heating analysis are proof-of-concept and only accounts for heating at entry and exit from the Roche-limit, we realistically expect a higher fraction of Par 1802-like systems to exhibit the observed temperature difference.

Finally, our results suggest that secular interaction in hierarchical triple body systems, during their pre-main-sequence contractions, may lead to a generic feature: tight binaries with temperature difference. Par 1802 is a young system ($\sim 2$~Myr), which results in limited time for the EKL mechanism to operate. We thus expect similar, more pervasive behaviour for general triple systems. In particular, the presence of a tertiary star may explain the $4$ other twin binaries with significant temperature differences discovered so far \citep{Gomez+09,Gomez+19,Gillen+2017,Wang+2009}. This dynamically-induced temperature difference may affect conclusions of the coevality of binary systems \citep[e.g.,][]{Kraus+09}.

\section*{Acknowledgements}
S.J.C. would like to acknowledge the support of the Lau Endowment as part of the Undergraduate Research Scholars Program of the UCLA Undergraduate Research Center - Sciences, the Undergraduate Research Fellows Program, and the Maggie Gilbert Research Award through UCLA College Honors.
S.N. thanks Howard and Astrid Preston for their generous support. We would like to thank Professor Stassun, Professor G\'{o}mez Maqueo Chew, Professor Heller, and Dr Hamers for their thoughtful comments and suggestions.
    
\bibliographystyle{mnras}
\bibliography{par18022}

\begin{thebibliography}{}
\makeatletter
\relax
\def\mn@urlcharsother{\let\do\@makeother \do\$\do\&\do\#\do\^\do\_\do\%\do\~}
\def\mn@doi{\begingroup\mn@urlcharsother \@ifnextchar [ {\mn@doi@}
  {\mn@doi@[]}}
\def\mn@doi@[#1]#2{\def\@tempa{#1}\ifx\@tempa\@empty \href
  {http://dx.doi.org/#2} {doi:#2}\else \href {http://dx.doi.org/#2} {#1}\fi
  \endgroup}
\def\mn@eprint#1#2{\mn@eprint@#1:#2::\@nil}
\def\mn@eprint@arXiv#1{\href {http://arxiv.org/abs/#1} {{\tt arXiv:#1}}}
\def\mn@eprint@dblp#1{\href {http://dblp.uni-trier.de/rec/bibtex/#1.xml}
  {dblp:#1}}
\def\mn@eprint@#1:#2:#3:#4\@nil{\def\@tempa {#1}\def\@tempb {#2}\def\@tempc
  {#3}\ifx \@tempc \@empty \let \@tempc \@tempb \let \@tempb \@tempa \fi \ifx
  \@tempb \@empty \def\@tempb {arXiv}\fi \@ifundefined
  {mn@eprint@\@tempb}{\@tempb:\@tempc}{\expandafter \expandafter \csname
  mn@eprint@\@tempb\endcsname \expandafter{\@tempc}}}

\bibitem[\protect\citeauthoryear{{Antonini}, {Murray}  \& {Mikkola}}{{Antonini}
  et~al.}{2014}]{Antonini+14}
{Antonini} F.,  {Murray} N.,   {Mikkola} S.,  2014, \mn@doi [The Astrophysical
  Journal] {10.1088/0004-637X/781/1/45}, \href
  {http://adsabs.harvard.edu/abs/2014ApJ...781...45A} {781, 45}

\bibitem[\protect\citeauthoryear{{Auclair-Desrotour}, {Mathis}  \& {Le
  Poncin-Lafitte}}{{Auclair-Desrotour} et~al.}{2015}]{Auclair+15}
{Auclair-Desrotour} P.,  {Mathis} S.,   {Le Poncin-Lafitte} C.,  2015, in
  European Physical Journal Web of Conferences. p. 04005,
  \mn@doi{10.1051/epjconf/201510104005}

\bibitem[\protect\citeauthoryear{{Baraffe}, {Chabrier}, {Allard}  \&
  {Hauschildt}}{{Baraffe} et~al.}{1998}]{Baraffe+98}
{Baraffe} I.,  {Chabrier} G.,  {Allard} F.,   {Hauschildt} P.~H.,  1998, \aap,
  \href {http://adsabs.harvard.edu/abs/1998A%26A...337..403B} {337, 403}

\bibitem[\protect\citeauthoryear{Bataille, Libert  \& Correia}{Bataille
  et~al.}{2018}]{Bataille+18}
Bataille M.,  Libert A.-S.,   Correia A. C.~M.,  2018, Monthly Notices of the
  Royal Astronomical Society, 479, 4749

\bibitem[\protect\citeauthoryear{{Bergfors} et~al.,}{{Bergfors}
  et~al.}{2010}]{Bergfors+10}
{Bergfors} C.,  et~al., 2010, \mn@doi [\aap] {10.1051/0004-6361/201014114},
  \href {https://ui.adsabs.harvard.edu/abs/2010A&A...520A..54B} {520, A54}

\bibitem[\protect\citeauthoryear{{Bode} \& {Wegg}}{{Bode} \&
  {Wegg}}{2014}]{BodeWegg+14}
{Bode} J.~N.,  {Wegg} C.,  2014, \mn@doi [\mnras] {10.1093/mnras/stt2227},
  \href {http://adsabs.harvard.edu/abs/2014MNRAS.438..573B} {438, 573}

\bibitem[\protect\citeauthoryear{{Braviner} \& {Ogilvie}}{{Braviner} \&
  {Ogilvie}}{2015}]{Braviner+15}
{Braviner} H.~J.,  {Ogilvie} G.~I.,  2015, \mn@doi [\mnras]
  {10.1093/mnras/stu2521}, \href
  {https://ui.adsabs.harvard.edu/abs/2015MNRAS.447.1141B} {447, 1141}

\bibitem[\protect\citeauthoryear{{Cargile}, {Stassun}  \& {Mathieu}}{{Cargile}
  et~al.}{2008}]{Cargile+08}
{Cargile} P.~A.,  {Stassun} K.~G.,   {Mathieu} R.~D.,  2008, \mn@doi [\apj]
  {10.1086/524346}, \href
  {https://ui.adsabs.harvard.edu/abs/2008ApJ...674..329C} {674, 329}

\bibitem[\protect\citeauthoryear{{Chabrier} \& {Baraffe}}{{Chabrier} \&
  {Baraffe}}{1997}]{Chabrier&Baraffe97}
{Chabrier} G.,  {Baraffe} I.,  1997, \aap, \href
  {https://ui.adsabs.harvard.edu/abs/1997A&A...327.1039C} {327, 1039}

\bibitem[\protect\citeauthoryear{{Dosopoulou} \& {Kalogera}}{{Dosopoulou} \&
  {Kalogera}}{2016}]{Dosopoulou+16}
{Dosopoulou} F.,  {Kalogera} V.,  2016, \mn@doi [\apj]
  {10.3847/0004-637X/825/1/71}, \href
  {https://ui.adsabs.harvard.edu/abs/2016ApJ...825...71D} {825, 71}

\bibitem[\protect\citeauthoryear{{Duquennoy} \& {Mayor}}{{Duquennoy} \&
  {Mayor}}{1991}]{Duquennoy+91}
{Duquennoy} A.,  {Mayor} M.,  1991, \aap, \href
  {https://ui.adsabs.harvard.edu/abs/1991A&A...248..485D} {500, 337}

\bibitem[\protect\citeauthoryear{{Eggleton}}{{Eggleton}}{1983}]{Eggleton1983}
{Eggleton} P.~P.,  1983, \mn@doi [The Astrophysical Journal] {10.1086/160960},
  \href {https://ui.adsabs.harvard.edu/\#abs/1983ApJ...268..368E} {268, 368}

\bibitem[\protect\citeauthoryear{{Eggleton} \& {Kiseleva-Eggleton}}{{Eggleton}
  \& {Kiseleva-Eggleton}}{2001}]{Eggleton2001}
{Eggleton} P.~P.,  {Kiseleva-Eggleton} L.,  2001, \mn@doi [The Astrophysical
  Journal] {10.1086/323843}, \href
  {http://adsabs.harvard.edu/abs/2001ApJ...562.1012E} {562, 1012}

\bibitem[\protect\citeauthoryear{{Eggleton}, {Kiseleva}  \& {Hut}}{{Eggleton}
  et~al.}{1998}]{Egg1998}
{Eggleton} P.~P.,  {Kiseleva} L.~G.,   {Hut} P.,  1998, \mn@doi [The
  Astrophysical Journal] {10.1086/305670}, \href
  {http://adsabs.harvard.edu/abs/1998ApJ...499..853E} {499, 853}

\bibitem[\protect\citeauthoryear{{Eggleton}, {Kisseleva-Eggleton}  \&
  {Dearborn}}{{Eggleton} et~al.}{2007}]{Egg+07}
{Eggleton} P.~P.,  {Kisseleva-Eggleton} L.,   {Dearborn} X.,  2007, in
  {Hartkopf} W.~I.,  {Harmanec} P.,   {Guinan} E.~F.,  eds,  IAU Symposium Vol.
  240, Binary Stars as Critical Tools Tests in Contemporary Astrophysics. pp
  347--355, \mn@doi{10.1017/S1743921307004280}

\bibitem[\protect\citeauthoryear{{Fabrycky} \& {Tremaine}}{{Fabrycky} \&
  {Tremaine}}{2007}]{FabryckyTremaine07}
{Fabrycky} D.,  {Tremaine} S.,  2007, \mn@doi [The Astrophysical Journal]
  {10.1086/521702}, \href
  {https://ui.adsabs.harvard.edu/#abs/2007ApJ...669.1298F} {669, 1298}

\bibitem[\protect\citeauthoryear{{Fabrycky}, {Johnson}  \&
  {Goodman}}{{Fabrycky} et~al.}{2007}]{Fabrycky+07}
{Fabrycky} D.~C.,  {Johnson} E.~T.,   {Goodman} J.,  2007, \mn@doi [The
  Astrophysical Journal] {10.1086/519075}, \href
  {https://ui.adsabs.harvard.edu/#abs/2007ApJ...665..754F} {665, 754}

\bibitem[\protect\citeauthoryear{{Ferraz-Mello}, {Rodr{\'\i}guez}  \&
  {Hussmann}}{{Ferraz-Mello} et~al.}{2008}]{Ferraz+08}
{Ferraz-Mello} S.,  {Rodr{\'\i}guez} A.,   {Hussmann} H.,  2008, \mn@doi
  [Celestial Mechanics and Dynamical Astronomy] {10.1007/s10569-008-9133-x},
  \href {https://ui.adsabs.harvard.edu/abs/2008CeMDA.101..171F} {101, 171}

\bibitem[\protect\citeauthoryear{{Fischer} \& {Marcy}}{{Fischer} \&
  {Marcy}}{1992}]{Fischer+92}
{Fischer} D.~A.,  {Marcy} G.~W.,  1992, \mn@doi [\apj] {10.1086/171708}, \href
  {https://ui.adsabs.harvard.edu/abs/1992ApJ...396..178F} {396, 178}

\bibitem[\protect\citeauthoryear{{Ford}, {Kozinsky}  \& {Rasio}}{{Ford}
  et~al.}{2000}]{Ford+2000}
{Ford} E.~B.,  {Kozinsky} B.,   {Rasio} F.~A.,  2000, \mn@doi [\apj]
  {10.1086/308815}, \href
  {https://ui.adsabs.harvard.edu/abs/2000ApJ...535..385F} {535, 385}

\bibitem[\protect\citeauthoryear{Gillen, Hillenbrand, David, Aigrain, Rebull,
  Stauffer, Cody  \& Queloz}{Gillen et~al.}{2017}]{Gillen+2017}
Gillen E.,  Hillenbrand L.~A.,  David T.~J.,  Aigrain S.,  Rebull L.,  Stauffer
  J.,  Cody A.~M.,   Queloz D.,  2017, \mn@doi [The Astrophysical Journal]
  {10.3847/1538-4357/aa84b3}, 849, 11

\bibitem[\protect\citeauthoryear{{Goldreich} \& {Soter}}{{Goldreich} \&
  {Soter}}{1966}]{Goldreich+66}
{Goldreich} P.,  {Soter} S.,  1966, \mn@doi [\icarus]
  {10.1016/0019-1035(66)90051-0}, \href
  {https://ui.adsabs.harvard.edu/abs/1966Icar....5..375G} {5, 375}

\bibitem[\protect\citeauthoryear{{G{\'o}mez Maqueo Chew}, {Stassun},
  {Pr{\v{s}}a}  \& {Mathieu}}{{G{\'o}mez Maqueo Chew} et~al.}{2009}]{Gomez+09}
{G{\'o}mez Maqueo Chew} Y.,  {Stassun} K.~G.,  {Pr{\v{s}}a} A.,   {Mathieu}
  R.~D.,  2009, \mn@doi [\apj] {10.1088/0004-637X/699/2/1196}, \href
  {https://ui.adsabs.harvard.edu/abs/2009ApJ...699.1196G} {699, 1196}

\bibitem[\protect\citeauthoryear{{G{\'o}mez Maqueo Chew}, {Stassun},
  {Pr{\v{s}}a}, {Stempels}, {Hebb}, {Barnes}, {Heller}  \&
  {Mathieu}}{{G{\'o}mez Maqueo Chew} et~al.}{2012}]{Gomez+12}
{G{\'o}mez Maqueo Chew} Y.,  {Stassun} K.~G.,  {Pr{\v{s}}a} A.,  {Stempels} E.,
   {Hebb} L.,  {Barnes} R.,  {Heller} R.,   {Mathieu} R.~D.,  2012, \mn@doi
  [\apj] {10.1088/0004-637X/745/1/58}, \href
  {https://ui.adsabs.harvard.edu/abs/2012ApJ...745...58G} {745, 58}

\bibitem[\protect\citeauthoryear{{G{\'o}mez Maqueo Chew} et~al.,}{{G{\'o}mez
  Maqueo Chew} et~al.}{2019}]{Gomez+19}
{G{\'o}mez Maqueo Chew} Y.,  et~al., 2019, \mn@doi [\aap]
  {10.1051/0004-6361/201833299}, \href
  {https://ui.adsabs.harvard.edu/abs/2019A&A...623A..23G} {623, A23}

\bibitem[\protect\citeauthoryear{{Greenberg}}{{Greenberg}}{2009}]{Greenberg+09}
{Greenberg} R.,  2009, \mn@doi [\apjl] {10.1088/0004-637X/698/1/L42}, \href
  {https://ui.adsabs.harvard.edu/abs/2009ApJ...698L..42G} {698, L42}

\bibitem[\protect\citeauthoryear{{Griffin}}{{Griffin}}{2012}]{Griffin2012}
{Griffin} R.~F.,  2012, \mn@doi [Journal of Astrophysics and Astronomy]
  {10.1007/s12036-012-9137-5}, \href
  {http://adsabs.harvard.edu/abs/2012JApA...33...29G} {33, 29}

\bibitem[\protect\citeauthoryear{Guillochon, Ramirez-Ruiz  \& Lin}{Guillochon
  et~al.}{2011}]{Guillochon+11}
Guillochon J.,  Ramirez-Ruiz E.,   Lin D.,  2011, \mn@doi [The Astrophysical
  Journal] {10.1088/0004-637x/732/2/74}, 732, 74

\bibitem[\protect\citeauthoryear{{Hamers} \& {Dosopoulou}}{{Hamers} \&
  {Dosopoulou}}{2019}]{Hamers+19}
{Hamers} A.~S.,  {Dosopoulou} F.,  2019, \mn@doi [\apj]
  {10.3847/1538-4357/ab001d}, \href
  {https://ui.adsabs.harvard.edu/abs/2019ApJ...872..119H} {872, 119}

\bibitem[\protect\citeauthoryear{{Heller}}{{Heller}}{2018}]{Heller+18}
{Heller} R.,  2018, arXiv e-prints, \href
  {https://ui.adsabs.harvard.edu/abs/2018arXiv180606601H} {p. arXiv:1806.06601}

\bibitem[\protect\citeauthoryear{{Heller}, {Jackson}, {Barnes}, {Greenberg}  \&
  {Homeier}}{{Heller} et~al.}{2010}]{Heller+10}
{Heller} R.,  {Jackson} B.,  {Barnes} R.,  {Greenberg} R.,   {Homeier} D.,
  2010, \mn@doi [\aap] {10.1051/0004-6361/200912826}, \href
  {http://adsabs.harvard.edu/abs/2010A%26A...514A..22H} {514, A22}

\bibitem[\protect\citeauthoryear{{Henry}, {Jao}, {Subasavage}, {Beaulieu},
  {Ianna}, {Costa}  \& {M{\'e}ndez}}{{Henry} et~al.}{2006}]{Henry+06}
{Henry} T.~J.,  {Jao} W.-C.,  {Subasavage} J.~P.,  {Beaulieu} T.~D.,  {Ianna}
  P.~A.,  {Costa} E.,   {M{\'e}ndez} R.~A.,  2006, \mn@doi [\aj]
  {10.1086/508233}, \href {http://adsabs.harvard.edu/abs/2006AJ....132.2360H}
  {132, 2360}

\bibitem[\protect\citeauthoryear{{Hut}}{{Hut}}{1980}]{Hut80}
{Hut} P.,  1980, \aap, \href
  {https://ui.adsabs.harvard.edu/abs/1980A&A....92..167H} {92, 167}

\bibitem[\protect\citeauthoryear{{Janson} et~al.,}{{Janson}
  et~al.}{2012}]{Janson+12}
{Janson} M.,  et~al., 2012, \mn@doi [\apj] {10.1088/0004-637X/754/1/44}, \href
  {https://ui.adsabs.harvard.edu/abs/2012ApJ...754...44J} {754, 44}

\bibitem[\protect\citeauthoryear{{Kiseleva}, {Eggleton}  \&
  {Mikkola}}{{Kiseleva} et~al.}{1998}]{Kiseleva1998}
{Kiseleva} L.~G.,  {Eggleton} P.~P.,   {Mikkola} S.,  1998, \mn@doi [\mnras]
  {10.1046/j.1365-8711.1998.01903.x}, \href
  {https://ui.adsabs.harvard.edu/#abs/1998MNRAS.300..292K} {300, 292}

\bibitem[\protect\citeauthoryear{{Kozai}}{{Kozai}}{1962}]{Kozai}
{Kozai} Y.,  1962, \mn@doi [\aj] {10.1086/108790}, \href
  {http://adsabs.harvard.edu/abs/1962AJ.....67..591K} {67, 591}

\bibitem[\protect\citeauthoryear{{Kraus} \& {Hillenbrand}}{{Kraus} \&
  {Hillenbrand}}{2009}]{Kraus+09}
{Kraus} A.~L.,  {Hillenbrand} L.~A.,  2009, \mn@doi [\apj]
  {10.1088/0004-637X/704/1/531}, \href
  {https://ui.adsabs.harvard.edu/abs/2009ApJ...704..531K} {704, 531}

\bibitem[\protect\citeauthoryear{{Lidov}}{{Lidov}}{1962}]{Lidov}
{Lidov} M.~L.,  1962, \mn@doi [\planss] {10.1016/0032-0633(62)90129-0}, \href
  {http://adsabs.harvard.edu/abs/1962P%26SS....9..719L} {9, 719}

\bibitem[\protect\citeauthoryear{{Liu}, {Guillochon}, {Lin}  \&
  {Ramirez-Ruiz}}{{Liu} et~al.}{2013}]{Liu+13}
{Liu} S.-F.,  {Guillochon} J.,  {Lin} D.~N.~C.,   {Ramirez-Ruiz} E.,  2013,
  \mn@doi [The Astrophysical Journal] {10.1088/0004-637X/762/1/37}, \href
  {http://adsabs.harvard.edu/abs/2013ApJ...762...37L} {762, 37}

\bibitem[\protect\citeauthoryear{{Mardling} \& {Aarseth}}{{Mardling} \&
  {Aarseth}}{2001}]{Mardling+01}
{Mardling} R.~A.,  {Aarseth} S.~J.,  2001, \mn@doi [\mnras]
  {10.1046/j.1365-8711.2001.03974.x}, \href
  {http://adsabs.harvard.edu/abs/2001MNRAS.321..398M} {321, 398}

\bibitem[\protect\citeauthoryear{{Mathis}}{{Mathis}}{2019}]{Mathis+19}
{Mathis} S.,  2019, in EAS Publications Series. pp 5--33,
  \mn@doi{10.1051/eas/1982002}

\bibitem[\protect\citeauthoryear{Moe \& Kratter}{Moe \&
  Kratter}{2018}]{Moe+2018}
Moe M.,  Kratter K.~M.,  2018, \mn@doi [The Astrophysical Journal]
  {10.3847/1538-4357/aaa6d2}, 854, 44

\bibitem[\protect\citeauthoryear{{Myll{\"a}ri}, {Valtonen}, {Pasechnik}  \&
  {Mikkola}}{{Myll{\"a}ri} et~al.}{2018}]{Myllari+18}
{Myll{\"a}ri} A.,  {Valtonen} M.,  {Pasechnik} A.,   {Mikkola} S.,  2018,
  \mn@doi [\mnras] {10.1093/mnras/sty237}, \href
  {http://adsabs.harvard.edu/abs/2018MNRAS.476..830M} {476, 830}

\bibitem[\protect\citeauthoryear{{Naoz}}{{Naoz}}{2016}]{Naoz16}
{Naoz} S.,  2016, \mn@doi [\araa] {10.1146/annurev-astro-081915-023315}, \href
  {http://adsabs.harvard.edu/abs/2016ARA%26A..54..441N} {54, 441}

\bibitem[\protect\citeauthoryear{Naoz \& Fabrycky}{Naoz \& Fabrycky}{2014}]{NF}
Naoz S.,  Fabrycky D.~C.,  2014, \mn@doi [The Astrophysical Journal]
  {10.1088/0004-637x/793/2/137}, 793, 137

\bibitem[\protect\citeauthoryear{{Naoz}, {Farr}, {Lithwick}, {Rasio}  \&
  {Teyssandier}}{{Naoz} et~al.}{2013a}]{Naoz+13sec}
{Naoz} S.,  {Farr} W.~M.,  {Lithwick} Y.,  {Rasio} F.~A.,   {Teyssandier} J.,
  2013a, \mn@doi [\mnras] {10.1093/mnras/stt302}, \href
  {http://adsabs.harvard.edu/abs/2013MNRAS.431.2155N} {431, 2155}

\bibitem[\protect\citeauthoryear{{Naoz}, {Kocsis}, {Loeb}  \& {Yunes}}{{Naoz}
  et~al.}{2013b}]{Naoz+13GR}
{Naoz} S.,  {Kocsis} B.,  {Loeb} A.,   {Yunes} N.,  2013b, \mn@doi [The
  Astrophysical Journal] {10.1088/0004-637X/773/2/187}, \href
  {http://adsabs.harvard.edu/abs/2013ApJ...773..187N} {773, 187}

\bibitem[\protect\citeauthoryear{{Naoz}, {Fragos}, {Geller}, {Stephan}  \&
  {Rasio}}{{Naoz} et~al.}{2016}]{Naoz+16}
{Naoz} S.,  {Fragos} T.,  {Geller} A.,  {Stephan} A.~P.,   {Rasio} F.~A.,
  2016, \mn@doi [\apjl] {10.3847/2041-8205/822/2/L24}, \href
  {http://adsabs.harvard.edu/abs/2016ApJ...822L..24N} {822, L24}

\bibitem[\protect\citeauthoryear{{Perets} \& {Fabrycky}}{{Perets} \&
  {Fabrycky}}{2009}]{PeretsFabrycky09}
{Perets} H.~B.,  {Fabrycky} D.~C.,  2009, \mn@doi [The Astrophysical Journal]
  {10.1088/0004-637X/697/2/1048}, \href
  {https://ui.adsabs.harvard.edu/#abs/2009ApJ...697.1048P} {697, 1048}

\bibitem[\protect\citeauthoryear{{Pribulla} \& {Rucinski}}{{Pribulla} \&
  {Rucinski}}{2006}]{Pri+06}
{Pribulla} T.,  {Rucinski} S.~M.,  2006, \mn@doi [\aj] {10.1086/503871}, \href
  {http://adsabs.harvard.edu/abs/2006AJ....131.2986P} {131, 2986}

\bibitem[\protect\citeauthoryear{{Raghavan} et~al.,}{{Raghavan}
  et~al.}{2010}]{Raghavan+10}
{Raghavan} D.,  et~al., 2010, \mn@doi [\apjs] {10.1088/0067-0049/190/1/1},
  \href {http://adsabs.harvard.edu/abs/2010ApJS..190....1R} {190, 1}

\bibitem[\protect\citeauthoryear{{Rappaport}, {Deck}, {Levine}, {Borkovits},
  {Carter}, {El Mellah}, {Sanchis-Ojeda}  \& {Kalomeni}}{{Rappaport}
  et~al.}{2013}]{Rappaport+13}
{Rappaport} S.,  {Deck} K.,  {Levine} A.,  {Borkovits} T.,  {Carter} J.,  {El
  Mellah} I.,  {Sanchis-Ojeda} R.,   {Kalomeni} B.,  2013, \mn@doi [The
  Astrophysical Journal] {10.1088/0004-637X/768/1/33}, \href
  {http://adsabs.harvard.edu/abs/2013ApJ...768...33R} {768, 33}

\bibitem[\protect\citeauthoryear{{Rose}, {Naoz}  \& {Geller}}{{Rose}
  et~al.}{2019}]{Rose+19}
{Rose} S.~C.,  {Naoz} S.,   {Geller} A.~M.,  2019, arXiv e-prints, \href
  {https://ui.adsabs.harvard.edu/abs/2019arXiv190312185R} {p. arXiv:1903.12185}

\bibitem[\protect\citeauthoryear{{Salpeter}}{{Salpeter}}{1955}]{Salpeter+55}
{Salpeter} E.~E.,  1955, \mn@doi [The Astrophysical Journal] {10.1086/145971},
  \href {http://adsabs.harvard.edu/abs/1955ApJ...121..161S} {121, 161}

\bibitem[\protect\citeauthoryear{{Sepinsky}, {Willems}, {Kalogera}  \&
  {Rasio}}{{Sepinsky} et~al.}{2007}]{Sepinsky+07}
{Sepinsky} J.~F.,  {Willems} B.,  {Kalogera} V.,   {Rasio} F.~A.,  2007,
  \mn@doi [\apj] {10.1086/520911}, \href
  {https://ui.adsabs.harvard.edu/abs/2007ApJ...667.1170S} {667, 1170}

\bibitem[\protect\citeauthoryear{{Shappee} \& {Thompson}}{{Shappee} \&
  {Thompson}}{2013}]{Shappee+13}
{Shappee} B.~J.,  {Thompson} T.~A.,  2013, \mn@doi [The Astrophysical Journal]
  {10.1088/0004-637X/766/1/64}, \href
  {http://adsabs.harvard.edu/abs/2013ApJ...766...64S} {766, 64}

\bibitem[\protect\citeauthoryear{{Stassun}, {Mathieu}, {Cargile}, {Aarnio},
  {Stempels}  \& {Geller}}{{Stassun} et~al.}{2008}]{Stassun+08}
{Stassun} K.~G.,  {Mathieu} R.~D.,  {Cargile} P.~A.,  {Aarnio} A.~N.,
  {Stempels} E.,   {Geller} A.,  2008, \mn@doi [\nat] {10.1038/nature07069},
  \href {https://ui.adsabs.harvard.edu/abs/2008Natur.453.1079S} {453, 1079}

\bibitem[\protect\citeauthoryear{{Stassun}, {Feiden}  \& {Torres}}{{Stassun}
  et~al.}{2014}]{Stassun+14}
{Stassun} K.~G.,  {Feiden} G.~A.,   {Torres} G.,  2014, \mn@doi [\nar]
  {10.1016/j.newar.2014.06.001}, \href
  {http://adsabs.harvard.edu/abs/2014NewAR..60....1S} {60, 1}

\bibitem[\protect\citeauthoryear{Stephan, Naoz, Ghez, Witzel, Sitarski, Do  \&
  Kocsis}{Stephan et~al.}{2016}]{Stephan+16}
Stephan A.~P.,  Naoz S.,  Ghez A.~M.,  Witzel G.,  Sitarski B.~N.,  Do T.,
  Kocsis B.,  2016, \mn@doi [Monthly Notices of the Royal Astronomical Society]
  {10.1093/mnras/stw1220}, 460, 3494

\bibitem[\protect\citeauthoryear{{Thompson}}{{Thompson}}{2011}]{Thompson2011}
{Thompson} T.~A.,  2011, \mn@doi [The Astrophysical Journal]
  {10.1088/0004-637X/741/2/82}, \href
  {https://ui.adsabs.harvard.edu/#abs/2011ApJ...741...82T} {741, 82}

\bibitem[\protect\citeauthoryear{{Tokovinin}}{{Tokovinin}}{1997}]{T97}
{Tokovinin} A.~A.,  1997, Astronomy Letters, \href
  {http://adsabs.harvard.edu/abs/1997AstL...23..727T} {23, 727}

\bibitem[\protect\citeauthoryear{{Tokovinin}, {Thomas}, {Sterzik}  \&
  {Udry}}{{Tokovinin} et~al.}{2006}]{Tok+06}
{Tokovinin} A.,  {Thomas} S.,  {Sterzik} M.,   {Udry} S.,  2006, \mn@doi [\aap]
  {10.1051/0004-6361:20054427}, \href
  {http://adsabs.harvard.edu/abs/2006A%26A...450..681T} {450, 681}

\bibitem[\protect\citeauthoryear{{Vick} \& {Lai}}{{Vick} \&
  {Lai}}{2018}]{Lai+18}
{Vick} M.,  {Lai} D.,  2018, \mn@doi [\mnras] {10.1093/mnras/sty225}, \href
  {https://ui.adsabs.harvard.edu/abs/2018MNRAS.476..482V} {476, 482}

\bibitem[\protect\citeauthoryear{{Wang}, {Wei}, {Shi}  \& {Zhao}}{{Wang}
  et~al.}{2009}]{Wang+2009}
{Wang} H.~J.,  {Wei} J.~Y.,  {Shi} J.~R.,   {Zhao} J.~K.,  2009, \mn@doi [\aap]
  {10.1051/0004-6361/200810873}, \href
  {https://ui.adsabs.harvard.edu/abs/2009A&A...500.1215W} {500, 1215}

\bibitem[\protect\citeauthoryear{{Wisdom}}{{Wisdom}}{2008}]{Wisdom+08}
{Wisdom} J.,  2008, \mn@doi [\icarus] {10.1016/j.icarus.2007.09.002}, \href
  {https://ui.adsabs.harvard.edu/abs/2008Icar..193..637W} {193, 637}

\bibitem[\protect\citeauthoryear{{Witte} \& {Savonije}}{{Witte} \&
  {Savonije}}{2002}]{Witte+02}
{Witte} M.~G.,  {Savonije} G.~J.,  2002, \mn@doi [\aap]
  {10.1051/0004-6361:20020155}, \href
  {https://ui.adsabs.harvard.edu/abs/2002A&A...386..222W} {386, 222}

\makeatother
\end{thebibliography}

% Don't change these lines
\bsp	% typesetting comment
\label{lastpage}
\end{document}